\documentclass{ws-ijmpd}

\begin{document}

\markboth{B.J. Ahmedov, A.V. Khugaev and N.I. Rakhmatov}
{Electromagnetic Fields of Charged and Magnetized  Cylindrical
Conductors in NUT Space    }

%
\catchline{}{}{}{}{}
%

\title{Electromagnetic Fields of Charged and
Magnetized \\ Cylindrical Conductors in NUT Space}
\author{\footnotesize B.J. AHMEDOV}

\address{Institute of Nuclear Physics and
    Ulugh Beg Astronomical Institute\\ Astronomicheskaya 33,
    Tashkent 700052, Uzbekistan\\
   IUCAA, Post Bag 4 Ganeshkhind,
411007 Pune, India\\
    ahmedov@astrin.uzsci.net}

\author{A.V. KHUGAEV}

\address{ Institute of Nuclear Physics,
         Ulughbek, Tashkent 702132, Uzbekistan         \\
         IUCAA,
         Post Bag 4 Ganeshkhind,
411007 Pune, India\\
avaskhugaev@yahoo.com }

\author{N.I. RAKHMATOV}

\address{ Institute of Nuclear Physics,
         Ulughbek, Tashkent 702132, Uzbekistan
         }

\maketitle

\begin{history}
\received{15 January 2004}
\end{history}

\begin{abstract}
We present analytic solutions of Maxwell equations for infinitely
long cylindrical conductors with nonvanishing electric charge and
currents in the external background spacetime of a line
gravitomagnetic monopole. It has been shown that vertical magnetic
field arising around cylindrical conducting shell carrying azimuthal
current will be
modified by the gravitational field of NUT source. We obtain that
the purely
general relativistic magnetic field which has no any Newtonian
analog will be produced around charged gravitomagnetic monopole.

\keywords{relativity stars; gravitomagnetic charge;
electromagnetic fields.}
\end{abstract}

{PACS Nos.: 04.20.-q; 04.20.Jb}

\section{Introduction}

General relativity predicts that so called gravitomagnetic
monopole or exotic NUT charge~\cite{nut63} creates a gravitomagnetic
field as in electrodynamics Dirac hypothetic monopole generates
magnetic field.
 In recent years, there exists an interest in problems
related to the general relativistic effects in the gravitational
field of NUT sources. For example, some interesting results
devoted to the physical aspects of the Kerr-NUT spaces have been
discussed in papers~\cite{bini03}$^{,}$ \cite{rh03}$^{,}$
\cite{nz02}. In particular in~\cite{bini03} Klein-Gordon and Dirac
equations in slowly rotating NUT spacetime have been investigated
and in~\cite{zonoz03} the solutions of the Maxwell equations for
the electromagnetic waves in Kerr-NUT space have been found.
In recent papers~\cite{rnz03},\cite{rh03} the theoretical effect of
magnetic mass in NUT space on the microlensing light curve and the
possibility of magnetic mass detection using the gravitational
microlensing technique have been studied.

In our previous research~\cite{99pla,00adp,03fp,ram01} the effect of
gravitomagnetic field of spinning gravitating body on the electric
current and electromagnetic field has been investigated on the
theoretical level. In fact, the influence of the angular momentum
of the rotating gravitational source may appear as a
galvanogravitomagnetic effect in the current carrying
conductors~\cite{99pla} and as a general-relativistic effect of
charge redistribution inside conductors in an applied magnetic
field ~\cite{00adp}. It has been shown in~\cite{03fp} that the
gravitomagnetic interaction with electric field can lead to the
applicable general-relativistic effects. The general relativistic
effect of ''dragging of inertial frames'' on electromagnetic
fields produced by rotating magnetic dipole has been investigated
in our papers~\cite{ram01}.

Here we extend our
research to the electromagnetic fields of conducting shell embedded in
the NUT space around cylindric gravitomagnetic monopole and find
analytical solutions of the Maxwell equations in this space-time.

The paper is organized as follows. In the section~\ref{meq} we
write the Maxwell equations in the exterior metric of cylindrical
NUT source. In the next section we solve Maxwell equations in NUT
space around i) conducting cylinder carrying electric current and
ii) electric charge put on line gravitomagnetic monopole. We
summarize our conclusions in the section~\ref{concl}.

Throughout, we use a space-like signature $(-,+,+,+)$ and a system
of units in which $G = 1 = c$ (However, for those expressions with
a physical application we have written the speed of light
explicitly.). Greek indices are taken to run from 0 to 3 and Latin
indices from 1 to 3; covariant derivatives are denoted with a
semi-colon and partial derivatives with a comma.

\section{Maxwell Equations In a Spacetime of Line Gravitomagnetic
Monopole}
\label{meq}

The main approximation in investigation of electromagnetic fields
in curved spacetime comes from neglecting the influence of the
electromagnetic field on the metric and by solving Maxwell
equations on a given, fixed background. Even for highly
magnetized compact stars the density of magnetic energy is
much smaller than the gravitational one. For example, if
$\rho$ is the density of a star of mass $M$ with radius $R$ the
ratio of the magnetic energy into the rest-mass energy
\begin{equation}
\frac{B^2}{8\pi\rho c^2}\simeq 1.6\times 10^{-6}
\left(\frac{B}{10^{15}G} \right)^2
\left(\frac{1.4 M_\odot}{M}\right)
\left(\frac{R}{15km} \right)^3 \
\end{equation}
is negligible small.

Electromagnetic energy of typical electric field of relativistic
compact stars
is also much less with compare to the energy of the gravitational field
\begin{equation}
\frac{E^2}{8\pi\rho c^2}\sim 0.2\times 10^{-20}
\left(\frac{E}{3\times 10^{10}V/cm} \right)^2
\left(\frac{1.4 M_\odot}{M}\right)
\left(\frac{R}{15km} \right)^3 \ .
\end{equation}

Our second approximation is in the specific
form of the background metric which we choose to be that of a
stationary, cylindrically symmetric system truncated at the first
order in gravitomagnetic monopole moment $L$. This approximation can
be justified by the fact that at the moment there is no any astrophysical
evidence for strong gravitomagnetic mass. For example, in the recent
paper~\cite{rh03} the magnetic mass
detection using the gravitational microlensing technique has been
explored and it has been evaluated that the minimum observable magnetic
mass to be about $14m$.

In a cylindrical coordinate system $(ct,r,\phi,z)$, the weak field
metric for external spacetime of a nonrotating cylindrical star
with nonvanishing gravitomagnetic charge is~\cite{nz97}
\begin{equation}
\label{weak_line}
ds^2 = -e^{-2\nu} dt^2 + e^{2\lambda}\left(dr^2
+ dz^2\right)+ r^2e^{2\nu} d\phi ^2 - 2 e^{-2\nu}Lz dt d\phi \ ,
\end{equation}
where the gravitomagnetic monopole momentum $L$, metric functions
$\lambda$ and $\nu$ are responsible for the gravitational field of
gravitomagnetic monopole.

According to the results of the above mentioned paper \cite{nz97},
we can extract dependence of functions $e^{-2\nu}$ and
$e^{2\lambda}$ in the following way (see Appendix for details on the
derivation of it)
 \begin{eqnarray}
 \label{2}
 && e^{-2\nu }=\frac{1}{(\frac{r}{c_0})^{2m}(\frac{L}{4m})^2+
 (\frac{r}{c_0})^{-2m}}\ , \\ \nonumber\\
 && e^{2\lambda }=r^{2m^2}[(\frac{r}{c_0})^{2m}(\frac{L}{4m})^2+
 (\frac{r}{c_0})^{-2m}]\ ,\\ \nonumber\\
\label{asymp2} && \lim_{L\to 0}e^{-2\nu}= \lim_{L\to
0}\frac{16m^2}{(\frac{\rho}{c_0})^{2m}L^2+
16m^2(\frac{\rho}{c_0})^{-2m}}=\nonumber \\ &&
\qquad\qquad\qquad\qquad\qquad\qquad
=(\frac{\rho}{c_0})^{2m}\Longrightarrow \lim_{m\to
0}(\frac{\rho}{c_0})^{2m}\to 1\ ,\\ \nonumber \\ && \lim_{L\to
0}e^{2\lambda}= \lim_{L\to
0}\frac{\rho^{2m^2}}{16m^2}[(\frac{\rho}{c_0})^{2m}L^2+
16m^2(\frac{\rho}{c_0})^{-2m}]=\nonumber \\ &&
\qquad\qquad\qquad\qquad\qquad
=\rho^{2m^2}(\frac{\rho}{c_0})^{-2m}\Longrightarrow
 \lim_{m\to 0}\rho^{2m^2}(\frac{\rho}{c_0})^{-2m}\to 1\ ,
\end{eqnarray}
where $c_0$ and $m$ are  integration constants~\footnote{$m$ is a
constant responsible for space-time curvature and for flat metric
 we choose $m=0$.}. The nonvanishing contravariant components of the metric
tensor are defined as
\begin{equation}
g^{00}=-e^{2\nu}\ , \qquad
g^{11}=g^{33}=e^{-2\lambda}\ , \qquad
g^{22}=\frac{e^{-2\nu}}{r^2}\ ,
\qquad
g^{02}=-\frac{e^{-2\nu}Lz}{r^2}\ .
\end{equation}

    The general form of the first pair of general
relativistic Maxwell equations is given by
\begin{equation}
\label{maxwell_firstpair}
3! F_{[\alpha \beta, \gamma]} =  2 \left(F_{\alpha \beta, \gamma }
    + F_{\gamma \alpha, \beta} + F_{\beta \gamma,\alpha}
    \right) = 0 \ ,
\end{equation}
where $F_{\alpha \beta}$ is the electromagnetic field tensor
expressing the strict connection between the electric and magnetic
four-vector fields $E^{\alpha},\ B^{\alpha}$. For an observer with
four-velocity $u^{\alpha}$, the covariant components of the
electromagnetic tensor are given by~\cite{l67}
\begin{equation}
\label{fab_def}
F_{\alpha\beta} \equiv 2 u_{[\alpha} E_{\beta]} +
    \eta_{\alpha\beta\gamma\delta}u^\gamma B^\delta \ ,
\end{equation}
where $T_{[\alpha \beta]} \equiv \frac{1}{2}(T_{\alpha
\beta} - T_{\beta \alpha})$ and
$\eta_{\alpha\beta\gamma\delta}$ is the pseudo-tensorial
expression for the Levi-Civita symbol $\epsilon_{\alpha
\beta \gamma \delta}$~\cite{s90}
\begin{equation}
\eta^{\alpha\beta\gamma\delta}=-\frac{1}{\sqrt{-g}}
    \epsilon_{\alpha\beta\gamma\delta} \ ,
    \hskip 2.0cm
\eta_{\alpha\beta\gamma\delta}=
    \sqrt{-g}\epsilon_{\alpha\beta\gamma\delta} \ ,
\end{equation}
with $g\equiv {\rm
det}|g_{\alpha\beta}|=- r^4
\sin^2\theta$ for the metric (\ref{weak_line}).

``Zero angular momentum observers'' or ZAMOs~\cite{bpt72} are locally
stationary (i.e. at fixed values of $r$) observers who
are ``dragged'' into rotation with respect to a reference frame
fixed with respect to distant observers. At first order in
gravitomagnetic monopole moment $L$
they have four-velocity components given by
\begin{equation}
\label{uzamos}
(u^{\alpha})_{_{\rm ZAMO}}\equiv
    e^{\nu}\bigg(1,0,0,
    \frac{e^{-4\nu}Lz}{r^2}\bigg) \ ;
    \hskip 2.0cm
(u_{\alpha})_{_{\rm ZAMO}}\equiv
    e^{-\nu}\bigg(- 1,0,0,0 \bigg) \ .
\end{equation}

    The general form of the second pair of Maxwell
equations is given by
\begin{equation}
\label{maxwell_secondpair} F^{\alpha \beta}_{\ \ \ \ ;\beta} =
4\pi J^{\alpha}\ ,
\end{equation}
where $J^{\alpha}$ is the four-current.

 It is important to project  Maxwell equations onto a locally orthonormal
tetrad because the 'hatted' components in an observer's basis define
measurable quantities (see, for example,~\cite{h04}).
Using (\ref{uzamos}) one can find  the components of
the tetrad $\{{\mathbf e}_{\hat \mu}\}$
%
\begin{eqnarray}
\label{zamo_tetrad_0} &&{\mathbf e}_{\hat 0}^{\alpha}  =
    e^{\nu}\bigg(1,0,\frac{e^{-3\nu}Lz}{r^2},0\bigg) \ ,    \\
\label{zamo_tetrad_1} &&{\mathbf e}_{\hat r}^{\alpha}  =
    e^{-\lambda}\bigg(0,1,0,0\bigg) \ ,     \\
\label{zamo_tetrad_2} &&{\mathbf e}_{\hat \phi}^{\alpha}  =
    \frac{e^{-\nu}}{r}\bigg(0,0,1,0\bigg)   \ ,         \\
\label{zamo_tetrad_3} &&{\mathbf e}_{\hat z}^{\alpha}  =
    e^{-\lambda}\bigg(0,0,0,1\bigg) \ ,
\end{eqnarray}
carried by ZAMO observer.

Following to~\cite{ram01} and with the definition
(\ref{fab_def}) referred to the observers (\ref{uzamos}) and ZAMO
orthonormal tetrad (\ref{zamo_tetrad_0})--(\ref{zamo_tetrad_3}),
we obtain Maxwell equations~(\ref{maxwell_firstpair}) and
(\ref{maxwell_secondpair}) in
useful form
\begin{eqnarray}
\label{max1a} && \left(re^{\lambda+\nu}B^{\hat r}\right)_{,r}+
    r^2e^{2(\lambda+\nu)}B^{\hat \phi}_{,\phi} +
    re^{\lambda+\nu} B^{\hat z}_{\ , \phi} = 0 \ ,
\\ \nonumber\\
\label{max1b} && re^{\lambda+\nu}\frac{\partial B^{\hat
r}}{\partial t}
    = rE^{\hat\phi}_{,z}- e^{\lambda-\nu}E^{\hat z}_{,\phi}
    -\frac{e^{\lambda-3\nu}Lz}{r} B^{\hat r}_{\ ,\phi} \ ,
\\ \nonumber\\
\label{max1c} && e^{2\lambda}\frac{\partial B^{\hat\phi}}{\partial
t}
    = \left( e^{\lambda-\nu} E^{\hat z}\right)_{,r}-
    e^{\lambda-\nu}E^{\hat r}_{,z}
    +\left(\frac{e^{\lambda-3\nu}Lz}{r} B^{\hat r}\right)_{\ ,r}
    +\left(\frac{e^{\lambda-3\nu}Lz}{r} B^{\hat z}\right)_{\ ,z}\
    ,
\\ \nonumber\\
\label{max1d} && re^{\lambda+\nu}\frac{\partial B^{\hat
z}}{\partial t}
    = -\left( rE^{\hat\phi}\right)_{,r}+
    e^{\lambda-\nu}E^{\hat r}_{,\phi}
    -\frac{e^{\lambda-3\nu}Lz}{r} B^{\hat z}_{\ ,\phi}\ ,
\end{eqnarray}
and
\begin{eqnarray}
\label{max2a}
&& \left(re^{\lambda+\nu} E^{\hat r} \right)_{,r}+
    e^{2\lambda}E^{\hat \phi}_{,\phi}
    + r e^{\lambda+\nu} E^{\hat z}_{\;,z}
     =  {4\pi re^{2\lambda+\nu}J^{\hat t}}\ ,
\\ \nonumber\\
\label{max2b} && \frac{e^{-\lambda-\nu}}{r} B^{\hat z}_{\;,\phi} -
e^{-2\lambda}B^{\hat \phi}_{\;,z}
    -\frac{e^{-3\nu-\lambda}Lz}{r^2}
    E^{\hat r}_{\ \;,\phi}
    = e^{\nu-\lambda}
    \frac{\partial E^{\hat r}}{\partial t}
    +4\pi e^{-\lambda} J^{\hat r}\ ,
\\ \nonumber\\
\label{max2c} && e^{\lambda-\nu}B^{\hat r}_{,z}-
\left(e^{\lambda-\nu} B^{\hat z} \right)_{,r}
    +\left(\frac{e^{\lambda-3\nu}Lz}{r}
    E^{\hat r}\right)_{\ \;,r}
    +\left(\frac{e^{\lambda-3\nu}Lz}{r}
    E^{\hat z}\right)_{\ \;,z}
    =\nonumber \\ \nonumber\\
&&\qquad\qquad\qquad\qquad = e^{2\lambda}
    \frac{\partial E^{\hat \phi}}{\partial t}
    +4\pi e^{2\lambda-\nu} J^{\hat \phi}+4\pi
    \frac{e^{2(\lambda-\nu)}Lz}{r}J^{\hat t} \ ,
\\ \nonumber\\
\label{max2d} && \left(r B^{\hat\phi} \right)_{,r} -
e^{\lambda-\nu}B^{\hat r}_{\;,\phi}
    -\frac{e^{\lambda-3\nu}Lz}{r}E^{\hat z}_{\ \;,\phi}
     = re^{\lambda+\nu}
    \frac{\partial E^{\hat z}}{\partial t} +
    4\pi re^{\lambda} J^{\hat z}\ .
\end{eqnarray}

\section{Stationary Vacuum Solutions to Maxwell Equations}
\label{ss}

    In this Section we will look for stationary
solutions of the Maxwell equation, i.e. for solutions in which we
assume that electromagnetic fields do not vary in time.

In the vacuum region all components of electric current are equal
to zero. Due to the stationarity and cylindrical symmetry of the
problem outside NUT source the electromagnetic field satisfies the
source-free Maxwell equations
\begin{equation}
\label{max1a_cyl}
\left(re^{\lambda+\nu}B^{\hat r}\right)_{,r}= 0 \ ,
\end{equation}
\begin{eqnarray}
\label{max1c_cyl}
&& \left( e^{\lambda-\nu} E^{\hat z}\right)_{,r}
    +\left(\frac{e^{\lambda-3\nu}Lz}{r} B^{\hat r}\right)_{\ ,r}
    =0 \ ,
\\
\label{max1d_cyl}
&&  \left( rE^{\hat\phi}\right)_{,r}=0\ ,
\end{eqnarray}

\noindent and
\begin{eqnarray}
\label{max2a_cyl}
&& \left(re^{\lambda+\nu} E^{\hat r} \right)_{,r}=0\ ,
\\
\label{max2c_cyl}
&& -\left(e^{\lambda-\nu} B^{\hat z} \right)_{,r}
    +\left(\frac{e^{\lambda-3\nu}Lz}{r}
    E^{\hat r}\right)_{\ \;,r}=0\ ,
\\
\label{max2d_cyl} &&\left(r B^{\hat\phi} \right)_{,r} =0\ ,
\end{eqnarray}
the solution of which are determined through the boundary
conditions at the surface of the sample.

Assume that the line NUT element is carrying electric current $I$
in radial direction. Then the azimuthal magnetic field in the
vicinity of the line element is given as
\begin{equation}
B^{\hat\varphi}=\frac{2 I}{r}
\end{equation}
and for this case there is no any modification of electromagnetic
field arising from the NUT charge.

We consider a simplified model of the source of electromagnetic
field made of conducting matter filling an infinitely long
cylindrical shell. Assume an infinitely long hollow cylindrical
($r_1$ and $r_2$ are radii of interior and exterior surfaces)
conductor carrying electric current in azimuthal direction ($I$ is
the electric current per unit length of the cylinder) is placed
symmetrically around the line NUT element. The vertical magnetic
field
\begin{equation}
\label{B_z}
B^{\hat z}=e^{-\lambda+\nu}4\pi I =4\pi I r^{-m^2}
\end{equation}
created by this current is modified by the gravitational field of
the NUT line source. Here we used the relations (\ref{2}). In the
limiting case when $m=0$ the magnetic field (\ref{B_z}) reduces to
its flat spacetime value.

It is naturally to investigate electromagnetic field generated by
the electric charge being put on the NUT source. Let's $Q$ is the
electric charge per unit length of the line tube. The external
electric field created by the electric charge is defined as
\begin{equation}
\label{ef}
E^{\hat r}=Q\frac{e^{-\lambda-\nu}}{r}=
\frac{16 m^2 Q r^{-m^2-1}}{(\frac{r}{c_0})^{2m}L^2+
16 m^2 (\frac{r}{c_0})^{-2m}}\ ,
\end{equation}
and the magnetic field is
\begin{eqnarray}
\label{mf}
&& B^{\hat{z}}=Qe^{-\lambda-3\nu}\frac{Lz}{r^2}= \frac{256 m^4 Q L z
r^{-m^2-2}}{[(\frac{r}{c_0})^{2m}L^2+ 16 m^2
(\frac{r}{c_0})^{-2m}]^2}\ .
\end{eqnarray}
The appearance of the magnetic field being proportional to the NUT
parameter $L$ is purely general relativistic result and has no any
Newtonian analog. At the limit $L\to 0$ one can see that magnetic
field (\ref{mf}) induced by the magnetic mass tends to zero. In
our case it corresponds to the metric with zero gravitomagnetic
charge.

One can see from figure 1 that electric field (\ref{ef}) behaviour
strongly
depends from gravitomagnetic charge density. In the limiting case
of flat spacetime when $m=0$ the electric field (\ref{ef}) reduces
to its flat spacetime Coulomb behaviour. Since there is no an
observational evidence for the existence of a gravitomagnetic mass,
the equations (\ref{ef}) and (\ref{mf}) could be in principle used
in order to detect the NUT parameter from analysis of the behaviour
of the electromagnetic fields around astrophysical objects.

\begin{figure}
\centerline{
\psfig{figure=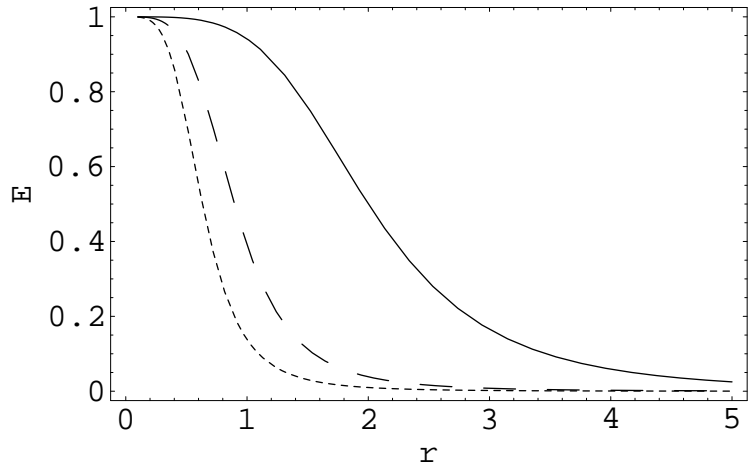,width=4.0in}}
\caption{Dependence of behaviour of electric field $E^{\hat r}(r)$
 from gravitomagnetic monopole momentum $L$. Solid line for $L=1$,
  dashed line for $L=5$ and dotted line for $L=10$. In all cases $m=1$.}
\end{figure}

\section{Conclusion}
\label{concl}

  We have presented analytic general relativistic
expressions for the electromagnetic fields external to the
cylindrical conductor either carrying electric current or having
electric charge in the space-time exterior to line gravitomagnetic
monopole which is considered isolated and in vacuum.

In the case of the carrying current conductor we have shown that
the magnetic field arising around it will be modified by the
gravitational field of NUT source. In the case of the charged
conductor an additional purely general relativistic magnetic field
will be induced by the magnetic mass.

In our future research  we will present
analytical solutions for the electromagnetic fields of slowly
rotating magnetized spherical NUT stars arising from the interplay
between gravitomagnetic charge and electric field.

\section*{Acknowledgments}

Financial support for this work is partially provided by the Abdus
Salam International Centre for Theoretical Physics through grant
AC-83 and by the NATO Reintegration Grant EAP.RIG.981259. Authors
thank TWAS for the support under the associate program and IUCAA
for warm hospitality during their stay in Pune. This research is
also supported in part by the UzFFR (project 01-06) and projects
F2.1.09, F2.2.06 and A13-226 of the UzCST.


\section*{Appendix: Comments to generalized cylindrical solution
of Einstein equation $\rho^2\nu^{''}+\rho\nu^{'}-
\frac{1}{2}\exp \left[-4\nu \right]L^2 = 0$}

\renewcommand{\theequation}{\Alph{equation}}
\setcounter{equation}{0}
Start from the Einstein vacuum equation for unknown metric function
$\nu(\rho)$ (equation (12a) from~\cite{nz97})
\begin{equation}
\label{equ1}
\rho^2\nu^{''}+\rho\nu^{'}-\frac{1}{2}e^{-4\nu }L^2 = 0\ ,
\end{equation}
which can be rewritten as
\begin{equation}
\label{equ2}
\rho\nu^{'} \left(\rho\nu^{'}\right)^{'}-
 \frac{1}{2}e^{-4\nu }\nu^{'}L^2 =0\ ,
\end{equation}
where $L$ is a constant. Equation~(\ref{equ2}) can be integrated as
\begin{equation}
\left(\rho\nu^{'}\right)^2+\frac{L^2}{4}e^{-4\nu}=c_1^2\ ,
\end{equation}
at $\nu^{'}\not= 0$, where  $c_1$ is the integration constant and
\begin{equation}
\label{equ3}
\int\frac{d\nu}{\sqrt{c_1^2-\frac{L^2}{4}e^{-4\nu}}}=
\pm ln\left(\frac{\rho}{c_2}\right)\ ,
\end{equation}
with the redefined integration constant $c_2$.
Introducing now new
variable $y=\sqrt{1-\left(\frac{L}{2c_1}e^{-2\nu}\right)^2}$
one can easily integrate
the expression~(\ref{equ3}) and
after making simple algebraic transformations obtain solution
of the Einstein equation~(\ref{equ1})
\begin{equation}
\label{equ4}
e^{-2\nu}=\frac{4c_1}{L}\frac{\left(\frac{\rho}{c_2}\right)^{2c_1}}
{\left(\frac{\rho}{c_2}\right)^{4c_1}+1}=
\frac{4c_1}{L}\frac{1}
{\left(\frac{\rho}{c_2}\right)^{2c_1}+
\left(\frac{\rho}{c_2}\right)^{-2c_1}}=
\frac{4m}{L}\frac{1}
{\left(\frac{\rho}{c}\right)^{2m}+
\left(\frac{\rho}{c}\right)^{-2m}}\ ,
\end{equation}
where we put $c_1=m$ and $c_2=c$. Thus the obtained solution is
identical to the results of Nouri-Zonos (see equations (13) and (14)
in the paper ~\cite{nz97}) if the integration constants
are slightly redefined for the correct description
of asymptotical properties of the solution. It is easy to
show that for the dimensional selfconsistence we have to select that
$\left[r\right]=\left[c\right]=\left[c_0\right]$ and
$\left[L\right]=\left[m\right]$.


\vspace*{6pt}

\begin{thebibliography}{0}
\bibitem{nut63}
E.T. Newman, L. Tamburino and T. Unti, {\it J. Math. Phys.} {\bf
4}, 915 (1963).

\bibitem{bini03}
{D. Bini, C. Cherubini, R.T. Jantzen and B. Mashhoon, {\it Class.
Quantum Grav.} {\bf 2}, 457 (2003); {\it Phys. Rev.} {\bf D67},
084013 (2003).}

\bibitem{zonoz03}{M. Nouri-Zonoz, {\it Class. Quantum Grav.}
{\bf 21}, 471 (2004).}

\bibitem{nz02}
{N. Dadhich and Z.Ya. Turakulov,  {\it Class. Quantum Grav.}{\bf
19}, 2765 (2002).}

\bibitem{rnz03}{S. Rahvar and M. Nouri-Zonoz,
{\it Mon. Not. R. Astron. Soc.}{\bf 338}, 926 (2003).}

\bibitem{rh03}{S. Rahvar and F. Habibi, {\it Ap. J.} {\bf 610},
673 (2004).}

\bibitem{99pla}
B.J. Ahmedov, {\it Phys. Lett.} {\bf A256}, 9 {(1999).}
\bibitem{00adp}
B.J. Ahmedov and M. Karim,  {\it Ann. Phys. (Leipzig)} {\bf 9},
SI-11 {(2000).}

\bibitem{ram01}{L. Rezzolla, B.J. Ahmedov and J.C. Miller,
{\it Mon. Not. R. Astron. Soc.}{\bf 322}, 723 (2001);
Erratum, {\it Mon. Not. R. Astron. Soc.}{\bf 338}, 816 (2003);
{\it Found. of Phys.}{\bf 31}, 1051 (2001).}

\bibitem{03fp} B.J. Ahmedov and N.I. Rakhmatov, {\it Found. Phys.}
{\bf 33}, 605 {(2003).}


\bibitem{nz97}{M. Nouri-Zonoz, {\it Class. Quantum Grav.} {\bf 14},
3123 (1997).}

\bibitem{l67}{A. Lichnerowicz, {\it Relativistic Hydrodynamics
    and Magnetohydrodynamics} (W.A. Benjamin, Inc, New York, 1967).}

\bibitem{s90}{H. Stephani,  {\it General Relativity}
(Cambridge Univ. Press., 1990).}

\bibitem{bpt72}{J.~M. Bardeen, W.H. Press, S.A. Teukolsky,
    {\it ApJ} {\bf 178}, 347 (1972).}



\bibitem{h04}{J.~B. Hartle, {\it Gravity. An Introduction to
Einstein's General Relativity} (Pearson Education, Inc, 2003).}

\end{thebibliography}
\end{document}